\documentclass[12pt,preprint]{aastex}

\usepackage{subfigure}

\def\msol{\rm{M}$_{\odot}$}
\def\lsol{\rm{L}$_{\odot}$}

\def\HII{H\,{\sc ii}}
\def\HI{H\,{\sc i}}
\def\arcsec{$^{\prime}$$^{\prime}$}

\def\deg{$^{\circ}$}
\def\micron{\,$\mu$m}
\def\etal{\textit{et al.}}

\def\an{AN}

\shortauthors{Mottram \etal{}}
\shorttitle{The Luminosity Functions and Timescales of MYSOs and Compact \HII{} regions}

\begin{document}

\title{The RMS Survey: The Luminosity Functions and Timescales of Massive Young Stellar Objects and Compact \HII{} Regions}

\author{Joseph~C.~Mottram\altaffilmark{1,2},
        Melvin~G.~Hoare\altaffilmark{2},
	Ben~Davies\altaffilmark{5,2},
        Stuart~L.~Lumsden\altaffilmark{2},
        Rene~D.~Oudmaijer\altaffilmark{2},
	James~S.~Urquhart\altaffilmark{3},
       Toby~J.~T.~Moore\altaffilmark{4},
        Heather~D.B.~Cooper\altaffilmark{2},
        Joseph~J.~Stead\altaffilmark{2}
}


\altaffiltext{1}{School of Physics, University of Exeter, Exeter, Devon, EX4 4QL, UK}

\altaffiltext{2}{School of Physics and Astronomy, University of Leeds, Leeds, LS2 9JT, UK}

\altaffiltext{3}{Australia Telescope National Facility, CSIRO Astronomy and Space Science, Sydney, NSW 2052, Australia}

\altaffiltext{4}{Astrophysics Research Institute, Liverpool John Moores University, Twelve Quays House, Egerton Wharf, Birkenhead, CH41 1LD, UK}

\altaffiltext{5}{Rochester Institute of Technology, 54 Lomb Memorial Drive, Rochester, NY 14623, USA}

\email{joe@astro.ex.ac.uk}

\begin{abstract}

We present a determination of the luminosity functions of massive young stellar objects (MYSOs) and compact (C) \HII{} regions within the Milky Way Galaxy using the large, well-selected sample of these sources identified by the Red MSX Source (RMS) survey. The MYSO luminosity function decreases monotonically such that there are few with $L\gtrsim$10$^{5}$\lsol{}, whilst the C\HII{} regions are detected up to $\sim$10$^{6}$\lsol{}.  The lifetimes of these phases are also calculated as a function of luminosity by comparison with the luminosity function for local main-sequence OB stars. These indicate that the MYSO phase has a duration ranging from 4$\times$10$^{5}$~yrs for 10$^{4}$\lsol{} to $\sim$7$\times$10$^{4}$~yrs at 10$^{5}$\lsol{}, whilst the C\HII{} region phase lasts of order 3$\times$10$^{5}$~yrs or $\sim$3-10$\%$ of the exciting star's main-sequence lifetime. MYSOs between 10$^{4}$\lsol{} and $\sim$10$^{5}$\lsol{} are massive but do not display the radio continuum or near-IR \HI{} recombination line emission indicative of an \HII{} region, consistent with being swollen due to high ongoing or recent accretion rates. Above $\sim$10$^{5}$\lsol{} the MYSO phase lifetime becomes comparable to the main-sequence Kelvin-Helmholtz timescale, at which point the central star can rapidly contract onto the main-sequence even if still accreting, and ionise a C\HII{} region, thus explaining why few highly luminous MYSOs are observed.

\end{abstract}


\section{Introduction}
\label{S:intro}

Massive stars (M~$\geq$~8\msol{}) dominate feedback processes in galaxies due to their strong ionising radiation, outflows, stellar winds and supernovae. They have the potential to both disrupt their natal molecular cloud and/or trigger further star formation around them. Regions of massive star formation are usually more heavily embedded within their parent molecular cloud and rarer than sites where only low mass stars form, thus we have to look further to find statistically significant samples. In addition, massive stars form and evolve faster than low mass stars \citep[e.g.][]{Shu1987}. These issues mean that there are few large, well-selected samples of young massive stars, making it difficult to see global trends over the variations of individual sources, and thus restricting our understanding of the formation and early evolution of massive stars. Original catalogues of mid-infrared bright but radio-quiet massive young stellar objects (MYSOs), where major accretion is probably still ongoing \citep[see e.g.][]{Palla1993,Behrend2001} but an \HII{} region has yet to form, contained only of order 50 sources \citep{Wynn-Williams1982,Henning1984}. MYSO samples based on IRAS colours have been constructed \citep[][]{Palla1991,Molinari1996,Sridharan2002} but are often hampered by confusion in regions of high source density due to the IRAS satellite's low spatial resolution. Large \HII{} regions were not included in those lists based on comparison with single dish radio surveys, but more compact \HII{} regions were found to contaminate these samples when followed up with higher resolution radio observations (Molinari et al. 1998). These criteria therefore tended to bias these samples away from complex regions where many massive stars form.

In order to alleviate this bias, we have conducted a survey to find MYSOs in the Galaxy. The Red MSX Source (RMS) survey is based on a colour-selected sample of $\sim$2000 candidate MYSOs by \citet{Lumsden2002} from the 18\arcsec{} resolution MSX mid-IR point source catalogue \citep[PSC,][]{Egan1999,Egan2003b}. The MSX PSC has a peak and mean source density of 1812~sources/deg$^{2}$ and 25~sources/deg$^{2}$ respectively, so source confusion is not a significant issue in the original sample selection. A series of complementary follow-up observations have been carried out designed to identify contaminants (low mass YSOs, evolved stars, proto-planetary nebulae and planetary nebulae) and provide the characteristics of the MYSOs, as well as the ultra-compact and compact \HII{} regions also present in the sample \citep[][]{Urquhart2008b}.  These have included $\sim$1\arcsec{} resolution radio continuum observations \citep{Urquhart2007a,Urquhart2009a}, $\sim$1\arcsec{} resolution mid-IR observations \citep[e.g.][]{Mottram2007} outside the SPITZER GLIMPSE survey regions \citep{Churchwell2009}, $^{13}$CO molecular line observations \citep{Urquhart2007c,Urquhart2008a,Urquhart2011} to obtain kinematic distances and low spectral-resolution near-IR spectroscopy \citep[e.g.][]{Clarke2006}. The spatial resolution of the sample is therefore not that of MSX (18\arcsec{}) but of the follow-up observations (1$-$2\arcsec{}). Source multiplicity within the MSX beam is easily revealed by the higher-resolution mid-IR follow-up \citep[$\sim$25$\%$][]{Mottram2007,Mottram2011}.

Each source within the RMS sample has been classified into one of several categories in the database at www.ast.leeds.ac.uk/RMS. `MYSO's are radio-quiet mid-IR point sources which are associated with $^{13}$CO emission. `\HII{} regions' are radio-loud and/or resolved in the mid-IR indicating emission from Ly$\alpha$ heated dust. The division between radio-loud \HII{} regions and radio-quiet MYSOs is clear from the plot of L$_{\rm rad}$$/$L$_{\rm bol}$ vs. L$_{\rm bol}$ in \citet{Hoare2007b}. Given the requirement to be an MSX point source ($\leq$18\arcsec{}), our \HII{} regions fall into the category of compact and ultra-compact \citep[e.g.][]{Habing1979}, although we will refer to them all as compact from now on for brevity. Compact \HII{} regions excited by stars with spectral types later than B1 may have radio surface brightnesses that are too low to be detected in our 1\arcsec{} radio observations, but would still exhibit strong hydrogen recombination emission with Case-B \citep[][]{Baker1938} line ratios in our near-IR spectroscopic observations (see \S\ref{S:conc}).

In \citet[][]{Mottram2011} we used far-IR fluxes from \citet[][]{Mottram2010} and other data, along with the spectral energy distribution (SED) model fitter of \citet{Robitaille2007a} to obtain luminosity measurements for $\sim$1200 of these sources. In this paper we use the luminosity distributions obtained by \citet[][]{Mottram2011} to calculate the luminosity functions for MYSOs and C\HII{} regions in the Galactic plane of our Galaxy. These calculations will be discussed in \S\ref{S:obtain_lfn}, the results of which will be presented in \S\ref{S:results}. We will then use these results to obtain estimates for the lifetime of these phases as a function of luminosity using a method similar to that of \citet{Wood1989b}, which will be discussed in \S\ref{S:timescales}. Our conclusions are then presented in \S\ref{S:conc}.

\section{The Luminosity Function}
\label{S:obtain_lfn}

\begin{figure}
\center
\subfigure{\includegraphics[width=0.49\textwidth]{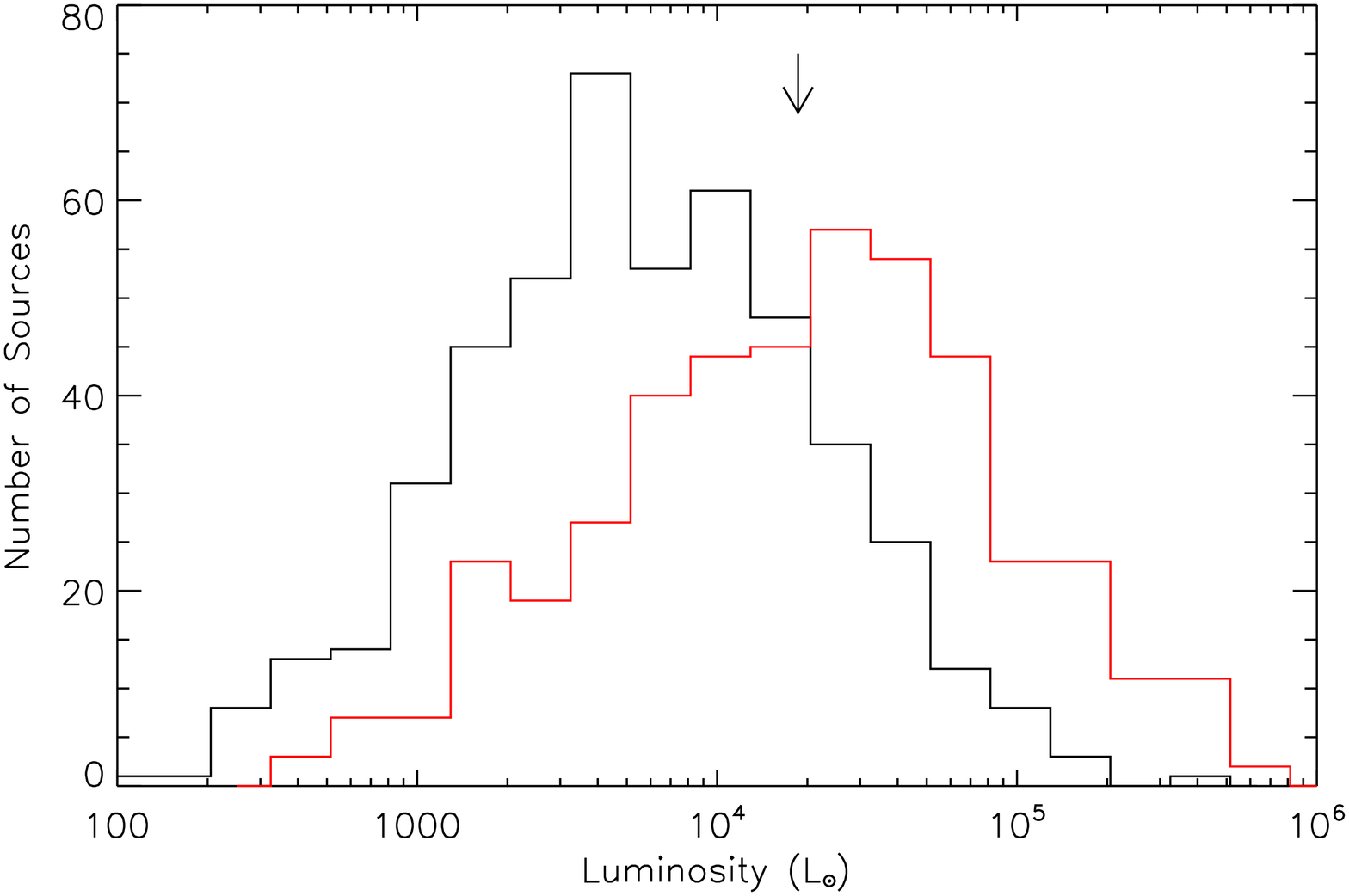}
\begin{picture}(10,1)(310,25)
\put(285,160){\makebox(0,0){(a)}}
\end{picture}
\label{F:obtain_lfn_input_a}}
\subfigure{\includegraphics[width=0.49\textwidth]{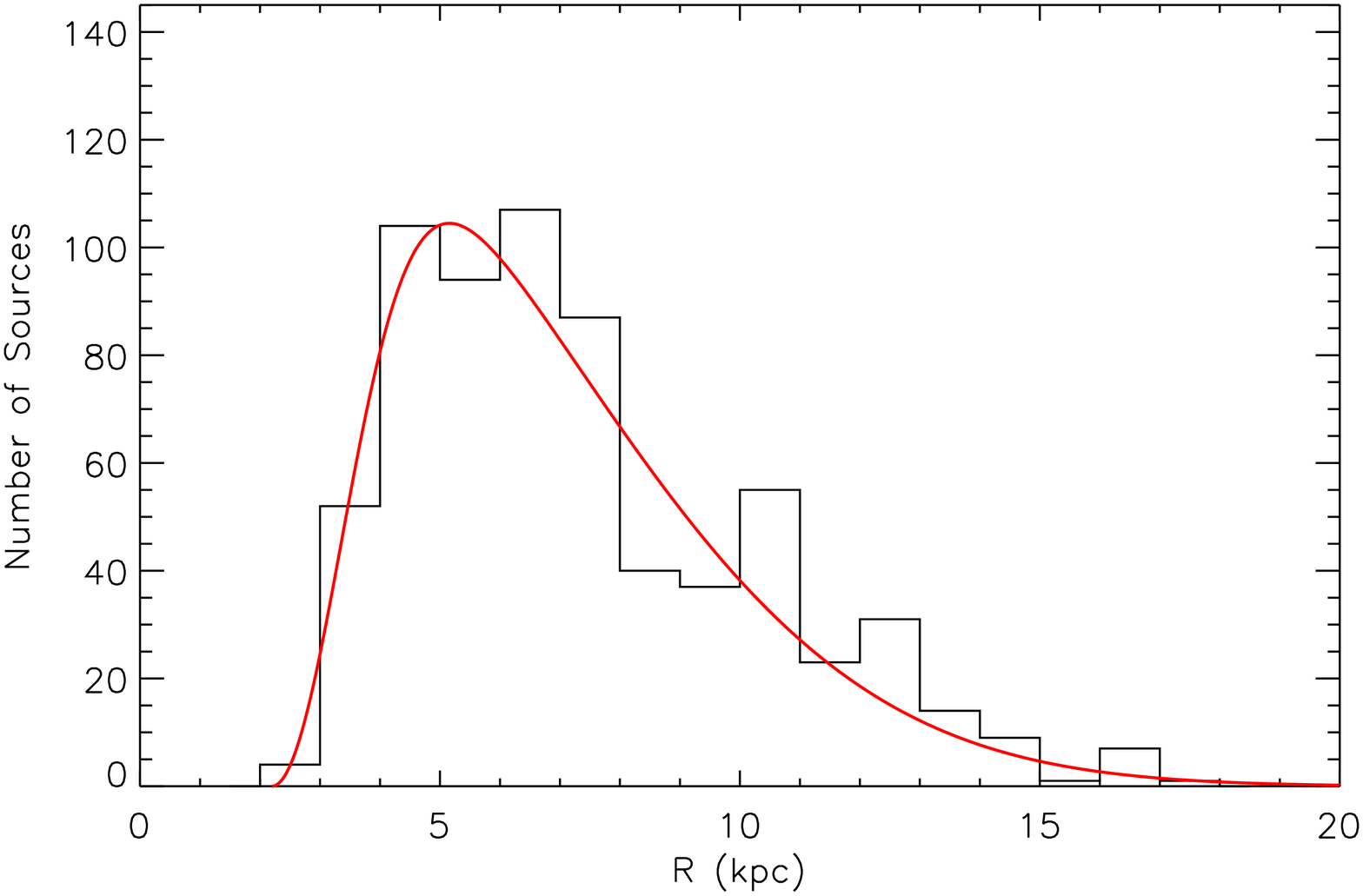}
\begin{picture}(10,1)(310,25)
\put(285,160){\makebox(0,0){(b)}}
\end{picture}
\label{F:obtain_lfn_input_b}}
\subfigure{\includegraphics[width=0.49\textwidth]{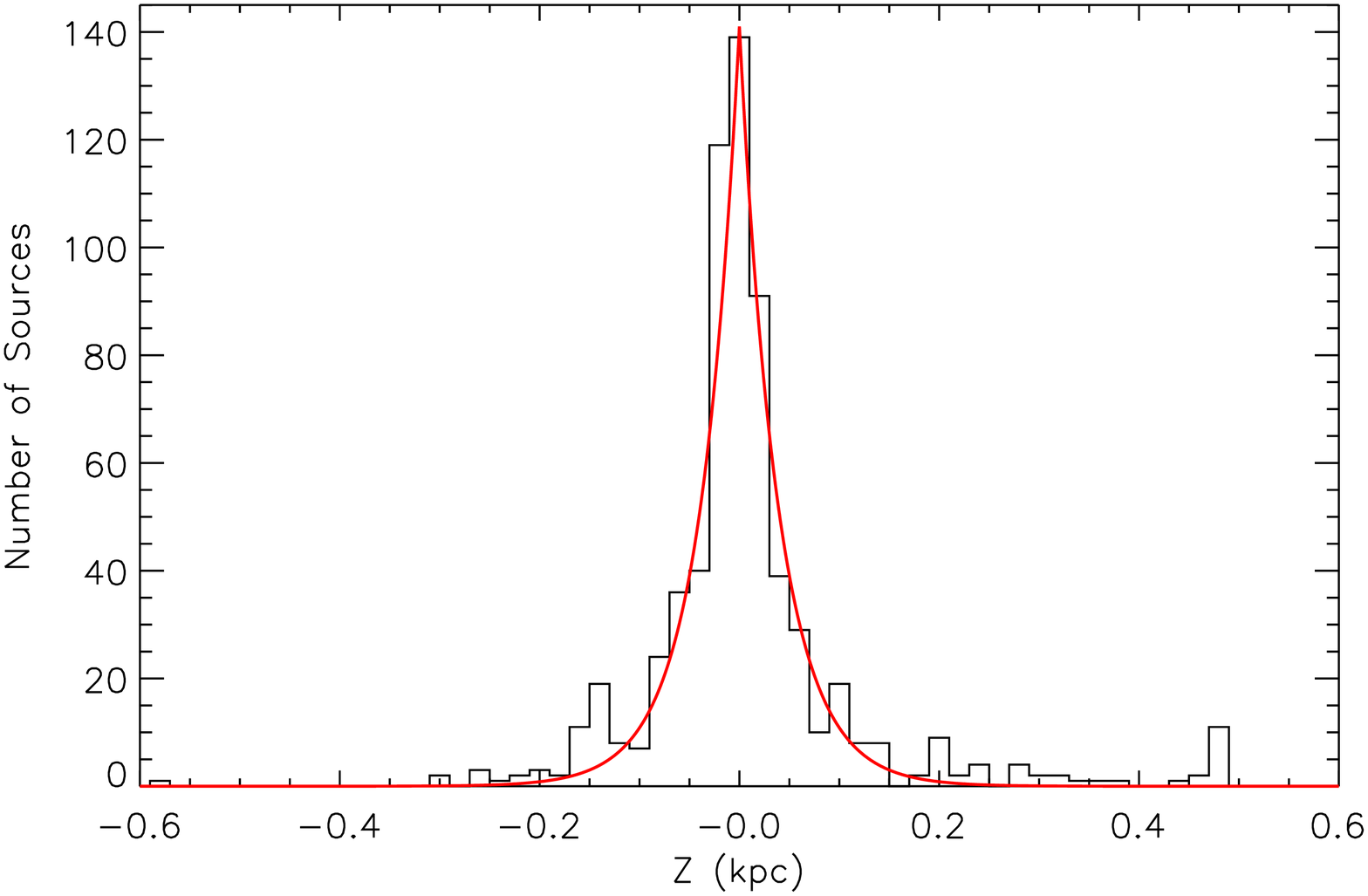}
\begin{picture}(10,1)(310,25)
\put(285,160){\makebox(0,0){(c)}}
\end{picture}
\label{F:obtain_lfn_input_c}}
\caption{Luminosity function inputs. (a): the number of sources vs. luminosity for RMS MYSOs (black) and C\HII{} regions (red) from \citet[][]{Mottram2011} rebinned to 0.2~dex. The black arrow indicate the luminosity of a B0V star from Table~\ref{T:obtain_lfn_spscales}. (b,c): the distribution of luminous ($L$~$\gtrsim$~8.2$\times$10$^{3}$\lsol{}) sources in the RMS survey as a function of galactocentric radius ($r$) and distance from the Galactic mid-plane ($z$) respectively. The red lines indicate the least-squares fit of the weighting function (equation~\ref{E:obtain_lfn_weight}) to the data.}
\label{F:obtain_lfn_input}
\end{figure}

We wish to derive a luminosity function defined as the number of objects per unit volume per unit luminosity interval in the Galaxy from the observed luminosity distributions, shown rebinned to 0.2 dex in Figure~\ref{F:obtain_lfn_input_a}. The first task is to examine the completeness of the sample as a function of luminosity. We do this by calculating the volume of the Galaxy within which MYSOs and C\HII{}s of luminosity L are detected by the RMS survey. A 3-D grid of cells was set up in cylindrical polar coordinates (i.e. $r$, $\theta$ and $z$), centred on the Galactic centre. The volume of each cell was calculated ($r$$\delta r$$\delta z$$\delta\theta$), along with the Galactic coordinates ($\ell$ $\&$ $b$) using the distance between the Earth and the Galactic centre of $R_{\rm 0}$~=~8.5~kpc \citep[][]{Marshall2006}. The distance between each cell and the Earth was also calculated, which was then used to obtain the total flux ($F_{\rm bol}$) that would be observed for a hypothetical source of luminosity $L$ within the cell.

Since the RMS sample is effectively a flux-limited sample in the MSX 21~\micron{} band we need to convert $F_{\rm bol}$ to $F_{\rm 21}$ for the hypothetical sources using the observed colours in the RMS sample. The observed distribution of $F_{\rm bol}$$/$$F_{\rm 21}$ was obtained from SED model fits for young RMS sources by \citet[][]{Mottram2011}, and takes the form of a Gaussian in log($F_{\rm bol}$$/$$F_{\rm 21}$) with mean 1.43 and standard deviation 0.27 \citep[see fig 7.,][]{Mottram2011}. It is well known the shape of the SEDs of C\HII{}s \citep{Chini1987} and MYSO \citep{Henning1990} are on average very similar, and no significant variation in $F_{\rm bol}$$/$$F_{\rm 21}$ between MYSOs and \HII{} regions was found in the model SEDs. The mean ratio was therefore used to calculate the MSX 21~\micron{} flux in Wm$^{-2}$ from $F_{\rm bol}$ for a hypothetical source of luminosity $L$ in the cell. These are then divided by the bandwidth of the MSX 21\micron{} band in order to convert the flux to Janskys. 

The completeness of the MSX survey at 21\micron{} varies with $\ell$ and $b$, particularly between the inner and outer galaxy due to the number of times each region of the sky was scanned, so we implement different flux limits ($F_{\rm limit}$) for these two regions. Given that most RMS sources lie in the Galactic mid-plane, we use $F_{\rm limit}$~=~3.5~Jy for 270\deg{}~$\geq$~$\ell$~$\geq$~90\deg{} and 4.25~Jy for 90\deg{}~$>$~$\ell$~$>$~270\deg{} based on the 50$\%$ completion limits calculated for $\mid$~$b$~$\mid$~$\leq$~1\deg{} by Davies \etal{} (2010, in prep). If a source with luminosity $L$ within the cell has $F_{\rm 21} \geq F_{\rm limit}$ then the volume of the cell is added to the total volume included in the RMS survey.

The observed distributions of young RMS sources with luminosities $L$~$\geq$~8.2$\times$10$^{3}$\lsol{} (discussed in \S\ref{S:results}) are shown in Figures \ref{F:obtain_lfn_input_b} and \ref{F:obtain_lfn_input_c} in terms of galactocentric radius ($r$) and distance from the Galactic mid-plane ($z$) respectively. Given that these are far from uniformly distributed within the Galaxy, a weighting function was used to weight the contribution made to the total volume by each cell according to the observed source distribution. This has the functional form of a thin disk with a hole in the centre as modelled by \citet{Robin2003} for the Galactic stellar population of the thin disk. We adopted length scales for the disk ($r_{\rm d}$) and central hole ($r_{\rm h}$) consistent with our observed distribution in galactocentric radius (see Figure~\ref{F:obtain_lfn_input_b}) and also allowed for an offset ($r_{\rm 0}$) from the Galactic centre to account for the observed lack of sources with $r \leq$ 2~kpc. The distribution of sources in terms of $z$ (Figure~\ref{F:obtain_lfn_input_c})  is that of an exponential with scale-height $z_{\rm d}$. The weight is therefore given by:

\begin{equation}
weight(r,z)~=~exp\left(\frac{-\mid z\mid}{z_{\rm d}}\right)~\left(exp\left(\frac{-(r-r_{\rm 0})^{2}}{r_{\rm d}^{2}}\right)-exp\left(\frac{-(r-r_{\rm 0})^{2}}{r_{\rm h}^{2}}\right)\right)
\label{E:obtain_lfn_weight}
\end{equation}

\noindent
A least-squares minimisation was used to obtain the values of the parameters in this function from the distribution of RMS sources, which resulted in $r_{\rm d}$~=~6.99~$\pm$~0.22~kpc, $r_{\rm h}$~=~1.71~$\pm$~0.23~kpc, $r_{\rm 0}$~=~2.20~$\pm$~0.10~kpc and $z_{\rm d}$~=~39.0~$\pm$~1.8~pc, shown by the red solid lines in Figures~\ref{F:obtain_lfn_input_b} and \ref{F:obtain_lfn_input_c}, with $\chi_{red}^{2}$=3.8 and 1.2 for the $r$ and $z$ fits respectively. This result does not change within the errors if luminosity cuts corresponding to 10$\%$ or 90$\%$ completeness are used instead of 50$\%$. The asymmetry in $z$ and the slight increase in the number of sources at $r$~=~10~kpc are probably due to local features near the Sun \citep[$R_{\rm 0}$~=~8.5~kpc, $Z_{\rm 0}$~=~15~pc,][]{Marshall2006}. Both the shape of the distributions and the fit results are similar to those obtained previously for sites of massive star formation by \citet{Bronfman2000} and for \HII{} regions by \citet{Paladini2004}.

The total weighted volume for which a source with luminosity $L$ should be included within the RMS survey is therefore calculated by integrating the weighted cell volume over all cells with $F_{\rm 21}~\geq F_{\rm limit}$, -5\deg{}~$\mid$~$b$~$\mid$~$\leq$~5\deg{} and 10\deg{}~$\geq$~$\ell$~$\geq$~350\deg{} \citep[see][]{Lumsden2002}. This process was then repeated for logarithmic steps in luminosity. Increments of 0.1~kpc, 0.5\deg{} and 0.01~kpc were used in $r$, $\theta$ and $z$ respectively, and all values were calculated for the cell centre.

The error in the luminosity is taken to be the bin width (0.2~dex) and the error in the luminosity function is calculated from statistical uncertainty using Poissonian statistics. Only bins with at least 5 sources in them are considered in the following analysis, ensuring that while the average uncertainty in source luminosity is $\sim$34$\%$ \citep{Mottram2011}, dominated by our conservative estimate of distance error, the ensemble uncertainty is smaller than the bin size.

\section{Results}
\label{S:results}

\begin{figure}
\center
\subfigure{\includegraphics[width=0.47\textwidth]{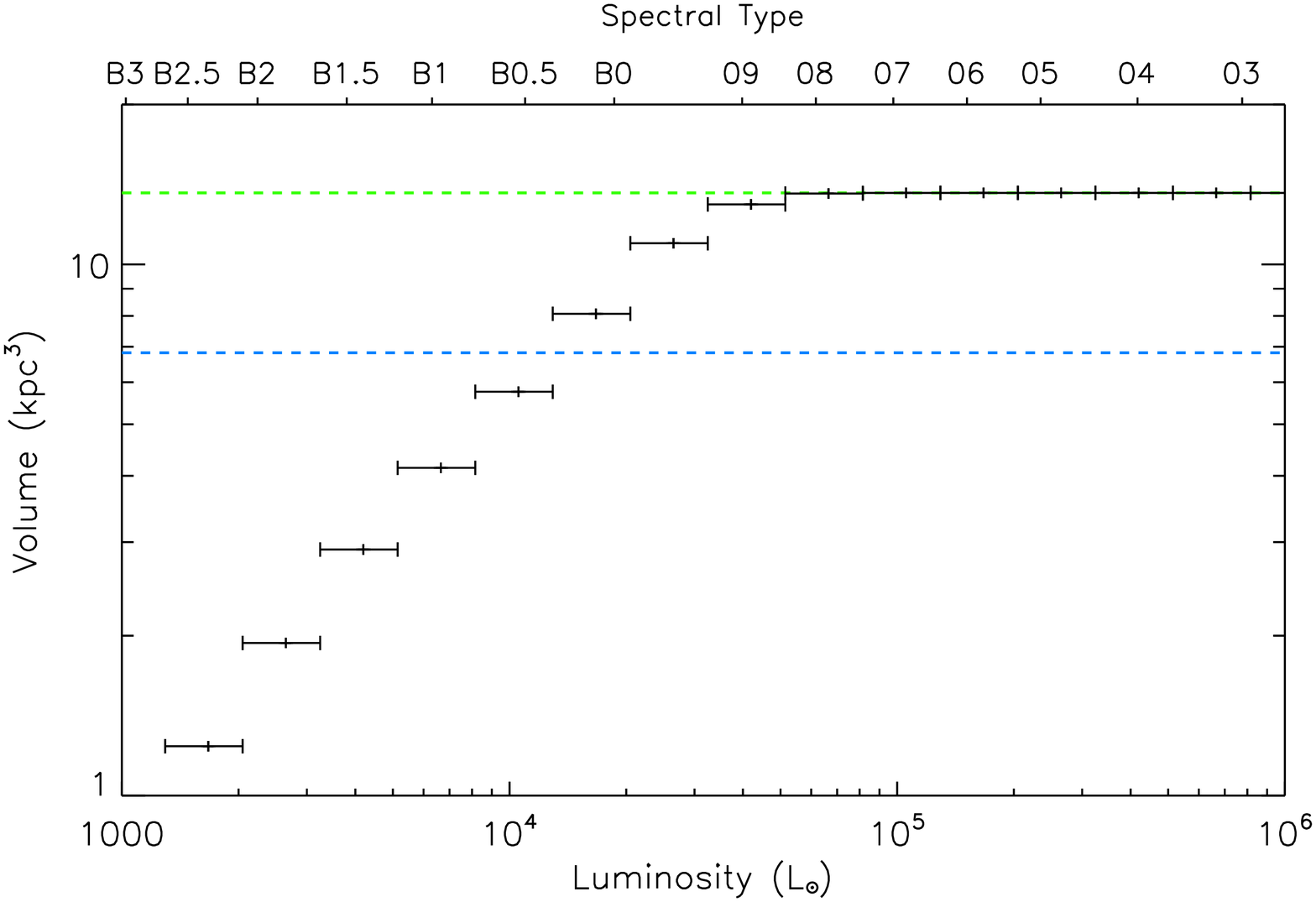}
\begin{picture}(10,1)(300,40)
\put(275,152){\makebox(0,0){(a)}}
\end{picture}
\label{F:results_lfuncs_a}}
\subfigure{\includegraphics[width=0.47\textwidth]{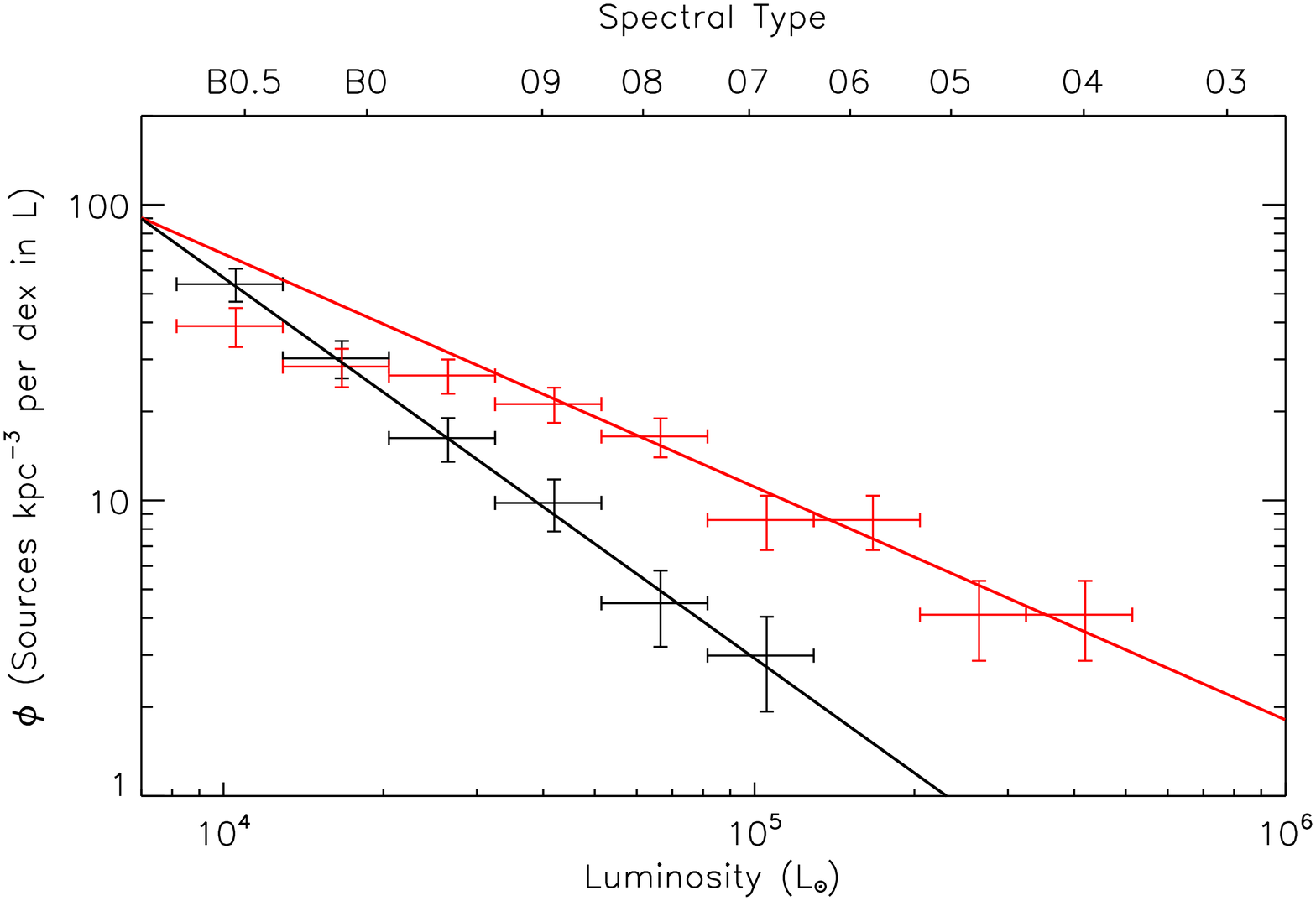}
\begin{picture}(10,1)(300,40)
\put(275,160){\makebox(0,0){(b)}}
\end{picture}
\label{F:results_lfuncs_b}}
\subfigure{\includegraphics[width=0.47\textwidth]{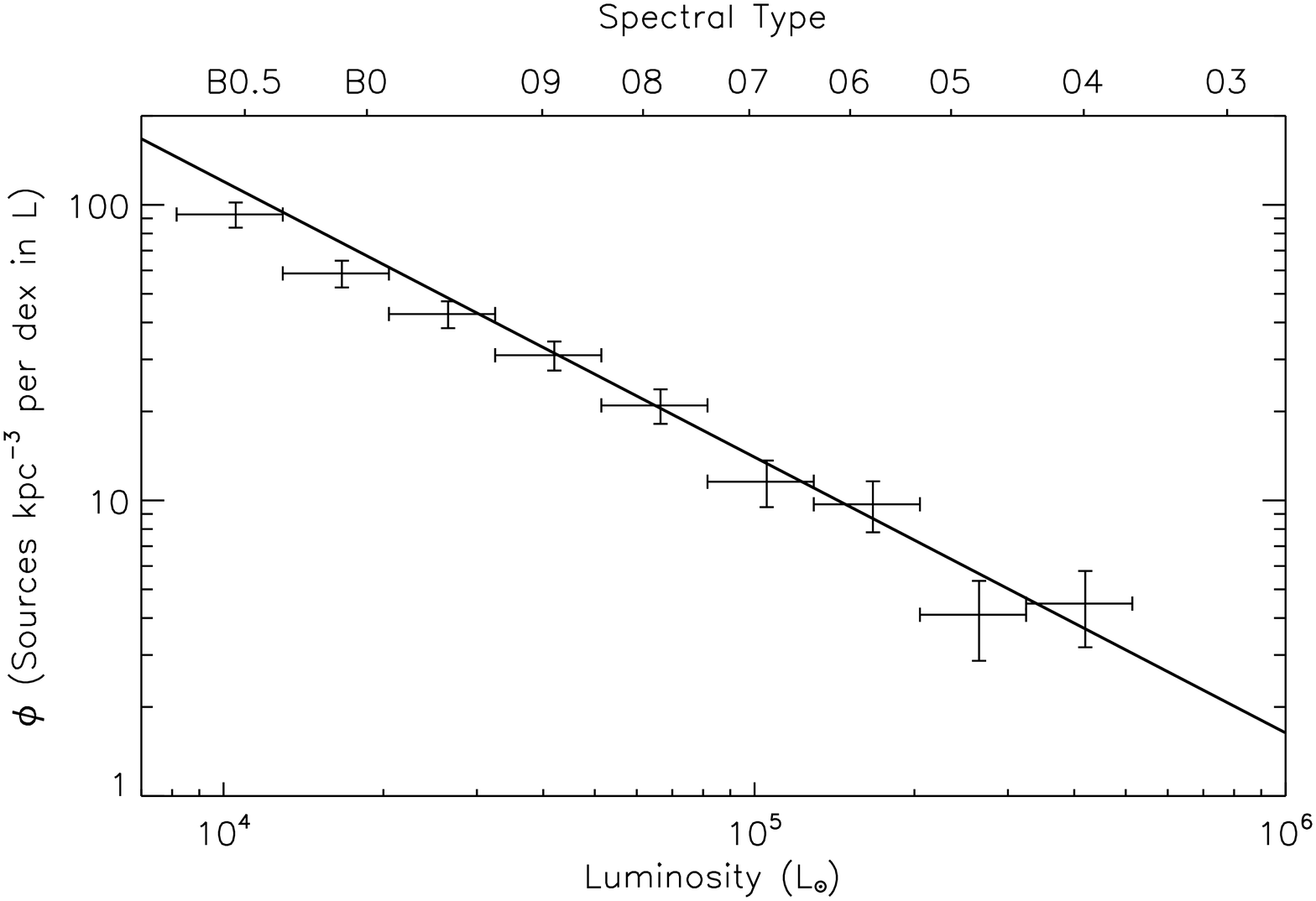}
\begin{picture}(10,1)(300,40)
\put(275,160){\makebox(0,0){(c)}}
\end{picture}
\label{F:results_lfuncs_c}}
\caption{Luminosity function results. (a): Volume of the Galaxy within which sources should be detected in the RMS survey vs. luminosity. The green and blue dashed lines indicate 100$\%$ (13.6~kpc$^{3}$) and 50$\%$ and completion respectively. (b): The luminosity functions for MYSOs (black) and C\HII{} regions (red). (c): The combined luminosity functions for sources identified as either MYSOs or C\HII{} regions in the RMS survey. The results of the total least-squares fits to the luminosity functions over limited ranges of $L$ are indicated by black and red solid lines, and are discussed in \S\ref{S:results}. The spectral type scale is presented in Table~\ref{T:obtain_lfn_spscales}}
\label{F:results_lfuncs}
\end{figure}

The volume of the Galaxy sampled by the RMS survey at each luminosity bin is shown in Figure~\ref{F:results_lfuncs_a}. This indicates that the RMS survey should be $\gtrsim$~50$\%$ complete above $\sim$8.2$\times$~10$^{3}$\lsol{}, which corresponds to a spectral type of $\sim$B0.5V using the scale presented in Table~\ref{T:obtain_lfn_spscales}. We restrict the source distribution calculations to above the 50$\%$ completion limit, as below this luminosity the volume correction required is $>$~2. The luminosity functions were then obtained by dividing the source numbers in Figure~\ref{F:obtain_lfn_input_a} by the volume in Figure~\ref{F:results_lfuncs_a} and converting to per dex in L. 

The resulting luminosity functions are shown above the 50$\%$ completion limit in Figure~\ref{F:results_lfuncs_b}. Over the luminosity range 10$^{4}$~\lsol{}~$\leq$~$L$~$\leq$~10$^{5}$~\lsol{}, the MYSO luminosity function falls steadily such that log($\phi_{\rm MYSO}$)~=~(-1.3~$\pm$~0.2)~log($L$/\lsol{}) + (6.9~$\pm$~0.9). Consequently, no significant population of MYSOs is observed above 10$^{5}$~\lsol{} or earlier than $\sim$O6.5V.

There is a flattening in the C\HII{} region luminosity function around $L$~=~4$\times$10$^{4}$~\lsol{} (09V).  Above this the luminosity function is fitted by log($\phi_{\rm CHII}$)~=~(-0.8~$\pm$~0.1)~log($L$/\lsol{}) + (5.0~$\pm$~0.8). Although the turnover in slope of the C\HII{} region luminosity function is below the 100$\%$ completeness limit, the fact that there is no similar break in the MYSO luminosity function that continues to lower luminosities means that the break is probably real.

Figure~\ref{F:results_lfuncs_c} shows the luminosity function for the combined sample of all MYSOs and C\HII{} regions identified in the RMS survey. This shows the overall function for the young mid-IR bright phase of massive star formation regardless of whether the source is ionising its surroundings or not. Above $L$~=~4$\times$10$^{4}$~\lsol{} this combined luminosity function is fitted by log($\phi_{\rm Combined}$)~=~(-0.9~$\pm$~0.2)~log($L$/\lsol{})~+~(5.8~$\pm$~0.8).

\section{Timescales}
\label{S:timescales}

\citet{Wood1989b} used the number of main-sequence stars to estimate the lifetime for all UC\HII{} regions, independent of luminosity, from the number of such sources they observed. The same method can be used to estimate the lifetimes of MYSOs and C\HII{} regions as a function of luminosity, using $\phi_{\rm MYSO}$($L$) and $\phi_{\rm CH\,{\sc II}}$($L$), i.e.:

\begin{equation}
t_{\rm MYSO}(L)~=~t_{\rm MS}(L)~\left(\frac{\phi_{\rm MYSO}(L)}{\phi_{\rm MS}(L)}\right)
\label{E:timescales_yso}
\end{equation}

\noindent where the main-sequence lifetime ($t_{\rm MS}$) of an OB star is given by log($t_{\rm MS}$/yrs)~=~(11.4~$\pm$~1.2)~-~(1.5~$\pm$~0.5)~log($L$/\lsol{})~+~(0.11~$\pm$~0.05)~log$^{2}$($L$/\lsol{}) calculated from a least-squares fit to the results of the models including stellar rotation by \citet{Meynet2003}. These models predict the correct ratio of red to blue supergiants, and are the only models which extend beyond 60~\msol{}. For M$\geq$20\msol{} we use the O star lifetime, while for M$<$20\msol{} we use the hydrogen-burning lifetime.
 
\begin{figure}
\center
\includegraphics[width=1.0\textwidth]{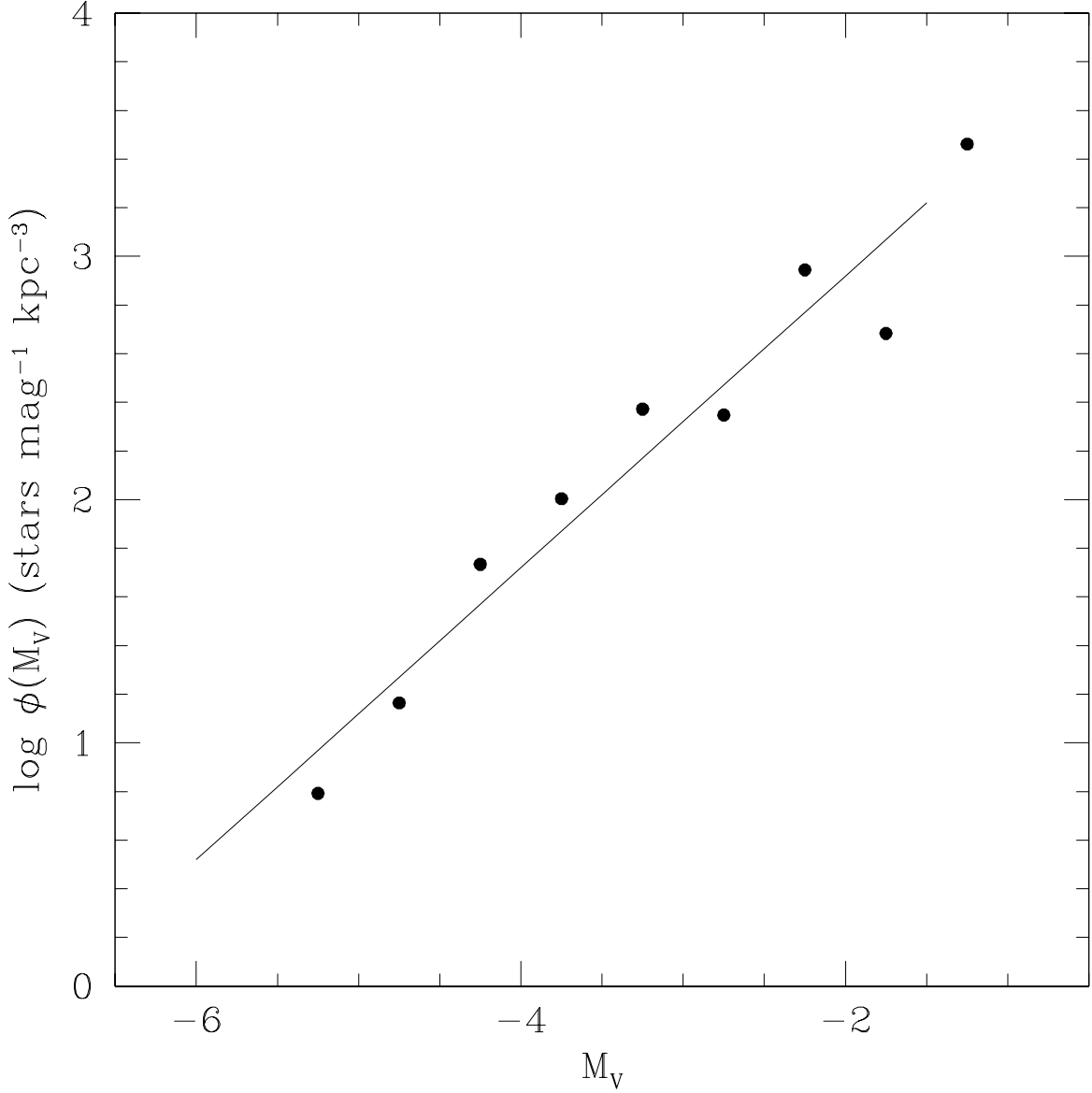}
\caption{Dwarf OB star luminosity function, calculated from the results of \citet[][]{Reed2005} using the $M_{\rm V}$-Spectral type scale given in Table~\ref{T:obtain_lfn_spscales}.}
\label{F:timescales_phiob}
\end{figure}

The local main-sequence OB star luminosity function in terms of $M_{\rm V}$ was obtained from re-binning the results on dwarf stars of \citet{Reed2005} corrected to the $M_{\rm V}$ scale in Table~\ref{T:obtain_lfn_spscales}, resulting in a least-squares fit of the form log($\phi_{\rm OB}(M_{\rm V})$/stars kpc$^{-3}$mag$^{-1}$)~=~(4.12~$\pm$~0.13) + (0.60~$\pm$~0.06)~$M_{\rm V}$. This is shown in Figure~\ref{F:timescales_phiob}.

This was then converted from $\phi_{\rm OB}(M_{\rm V})$ to $\phi_{\rm OB}(L)$ using $M_{\rm V}$~=~(1.5~$\pm$~0.3)~-~(0.6~$\pm$~0.1)~log($L$/\lsol{})~-~(0.10~$\pm$~0.02)~log$^{2}$($L$/\lsol{}) which is a least-squares fit to the data in Table~\ref{T:obtain_lfn_spscales}. We multiply our luminosity functions by the weight for the Galactic location of the Sun (i.e. weight($R_{\rm 0}$,$Z_{\rm 0}$)~=~0.302 from Eqn.~\ref{E:obtain_lfn_weight}) before using Eqn.~\ref{E:timescales_yso} so that they can be compared directly to the local OB luminosity function.

\begin{table}
\centering
\caption{Adopted properties of main-sequence stars. The surface temperatures are taken from \citet{Martins2005} and \citet{Boehm-Vitense1981} for O and B stars respectively. The luminosities and masses were calculated using a 1~Myr isochrone, chosen to better represent the observed local OB population, from the models with stellar rotation (v$_{\rm ini}$=300~kms$^{-1}$) of \citet[][]{Meynet2003}. $M_{\rm V}$ was calculated using the equations given in \citet{Martins2005} and \citet{Lanz2007} for O and B type stars respectively. The main-sequence lifetimes were calculated from the quadratic fit (see \S\ref{S:timescales}) to the main-sequence lifetimes of \citet{Meynet2003}. For a full discussion of the uncertainties in obtaining such data, see \citet{Martins2005}.}
\begin{tabular}{@{~}lcccccc@{~}}
\centering
&&&&&&\\
\hline
 Sp&T$_{\rm eff}$&log($L$)&$M$&$M_{\rm V}$&t$_{\rm KH}$&t$_{\rm MS}$\\
Type&(10$^{3}$~K)&(\lsol{})&(\msol{})&(mag)&(10$^{4}$~yrs)&(10$^{6}$~yrs)\\
\hline
O3&44.8&5.86&70.8&-5.9& 1.6& 2.9\\
O4&42.9&5.60&52.3&-5.3& 1.9& 3.3\\
O5&40.9&5.36&39.4&-4.9& 2.3& 3.9\\
O5.5&39.9&5.27&36.5&-4.7& 2.5& 4.2\\
O6&38.9&5.17&33.5&-4.6& 2.8& 4.5\\
O6.5&37.9&5.08&30.6&-4.4& 3.1& 4.9\\
O7&36.9&4.98&27.5&-4.2& 3.3& 5.4\\
O7.5&35.9&4.88&24.7&-4.1& 3.6& 5.9\\
O8&34.9&4.78&22.9&-3.9& 4.1& 6.5\\
O8.5&33.9&4.69&21.0&-3.7& 4.5& 7.2\\
O9&32.9&4.59&19.4&-3.6& 5.1& 8.1\\
O9.5&31.9&4.50&18.1&-3.4& 5.7& 8.9\\
B0&29.5&4.27&14.9&-3.1& 7.3&12.0\\
B0.5&27.3&4.04&12.7&-2.7&10.0&16.4\\
B1&25.0&3.80&10.7&-2.3&13.6&23.6\\
B1.5&23.0&3.58& 9.0&-1.9&17.4&33.8\\
B2&21.5&3.37& 7.9&-1.5&24.3&48.7\\
B2.5&20.2&3.18& 6.9&-1.2&31.5&69.2\\
B3&19.0&3.01& 6.0&-0.9&37.9&96.4\\
\hline
\end{tabular}
\label{T:obtain_lfn_spscales}
\end{table}

\begin{figure}
\center
\includegraphics[width=1.0\textwidth]{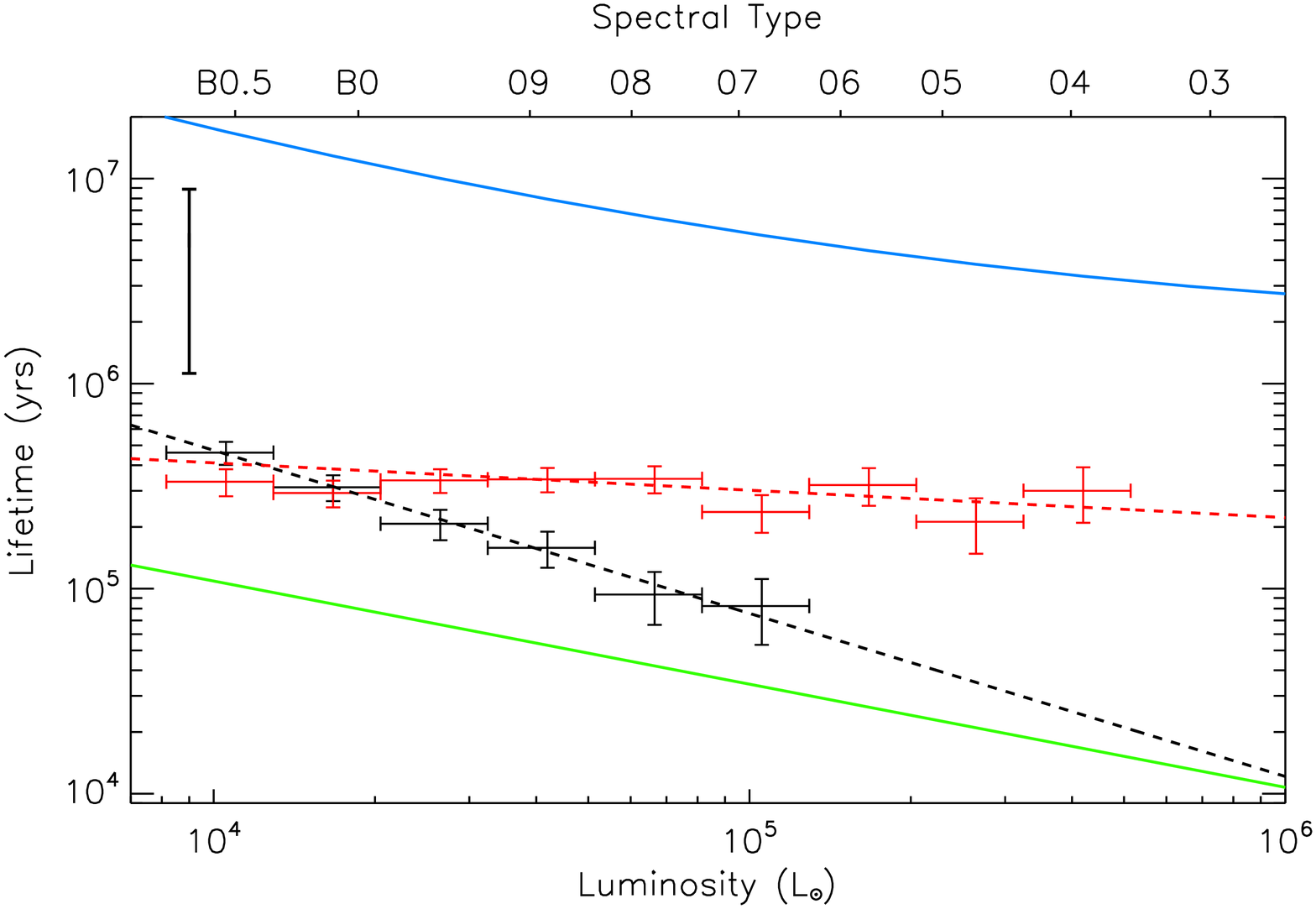}
\caption{The lifetime of MYSOs (black) and C\HII{} regions (red) vs. luminosity. The red and black dashed lines indicate the results of least-squares fits to the functions over limited ranges of $L$ and are discussed in \S\ref{S:timescales}. The green and blue solid lines indicate the main-sequence Kelvin-Helmholtz timescale and main-sequence lifetime respectively. The black vertical error bar in the top left corner indicates the systematic uncertainties.}
\label{F:timescales_yso}
\end{figure}

The MYSO and \HII{} region lifetimes obtained as functions of luminosity are shown in Figure~\ref{F:timescales_yso}. The difference between the fits to the data for $M_{\rm V}$ vs. log($L$), $\phi_{\rm OB}$ vs. $M_{\rm V}$ and log($t_{\rm MS}$) vs. log($L$) and the data themselves on the timescale calculations result in a systematic uncertainty of approximately a factor of 10, as shown by the black error bar. Fits to the data give log($t_{\rm MYSO}$/yrs)~=~(-0.80~$\pm$~0.24)~log($L$/\lsol{})~+~(8.9~$\pm$~1.1) and log($t_{\rm CHII}$/yrs)~=~(-0.13~$\pm$~0.16)~log($L$/\lsol{})~+~(6.1~$\pm$~0.8). The systematic error in the slope is of order 0.5, obtained by considering the errors in each step of the transformation. 

The timescale of MYSOs decreases by a little less than an order of magnitude beween 10$^{4}$\lsol{} and 10$^{5}$\lsol{}, while the C\HII{} region lifetime is flat with respect to luminosity at $\sim$3$\times$10$^{5}$yrs. \citet{Wood1989b} concluded that the lifetimes of \HII{} regions powered by O stars are about 15$\%$ of their main-sequence lifetimes. Here we find the C\HII{} region lifetimes are $\sim$3$-$10$\%$ of their main-sequence lifetimes, although the main-sequence lifetimes we are using are about twice as long as those used by \citet{Wood1989b} so our results are broadly consistent with theirs.

\section{Discussion \& Conclusions}
\label{S:conc}

We have obtained the luminosity functions of MYSOs and compact \HII{} regions in the Galactic plane. These clearly show that there is no significant population of the mid-IR bright, radio-quiet MYSOs above $\sim$10$^{5}$\lsol{}. This lack of high luminosity MYSOs could mean that the immediate progenitors of the most luminous C\HII{} regions are mid-IR faint or dark. Luminous far-IR cores and methanol masers that are too faint to be in our MSX-selected sample are being found \citep[e.g.][]{Chambers2009,Ellingsen2006}.

However, it is also important to consider the luminosity function in terms of the mechanism that must inhibit the formation of \HII{} regions around MYSOs, which are energetic enough to ionise their surroundings if they were on the main-sequence. Previously it has been suggested that the accretion rate and/or high density of the surrounding material could quench or trap the \HII{} region close to a young massive star \citep[e.g.][]{Walmsley1995,Keto2003,Tan2003}. In this case the \HII{} region would have a high enough emission measure (EM) to be optically thick in the radio continuum, e.g. $\tau_{\rm 1~cm}$~=~3$\times$10$^{-10}$~EM~(cm$^{-6}$pc)$\gg$1 \citep{Osterbrock1989}. Unless such regions have EM~$>$~10$^{16}$ near-IR recombination lines will still be optically thin since their line centre optical depths are much lower, e.g. $\tau_{\rm Br\gamma}$~$\sim$2$\times$10$^{-6}$~$\tau_{\rm 1~cm}$ \citep{Hummer1987}. This would lead to narrow lines with high equivalent widths whereas in MYSOs where we detect the NIR continuum, we actually observe broad lines with low equivalent widths consistent with optically thick emission from stellar winds \citep[e.g.][]{Bunn1995}.

Alternatively, as first suggested by \citet{Hoare2007b}, the MYSO could be too cool to ionise its surroundings on account of being swollen due to high accretion rates. Recent stellar evolution calculations including high accretion rates \citep{Yorke2008,Hosokawa2009,Hosokawa2010} have suggested that the radius of stars with masses between $\sim$8\msol{} and $\sim$30\msol{} increase dramatically (i.e. of order 10-100 times the main-sequence size) due to high accretion rates. Formation of an \HII{} region around an actively accreting star only begins when it is massive enough that the Kelvin-Helmholtz timescale is less than the accretion timescale. We have shown that the timescale associated with MYSOs becomes comparable with the main-sequence Kelvin Helmholtz timescale at $\sim$10$^{5}$\lsol{}. That we see no significant population of MYSOs with $L>$10$^{5}$\lsol{} is consistent with this idea. For the most massive MYSOs, i.e. those that have reached $\sim$30\msol{}, the accretion rate must be between 10$^{-4}$\msol{}~yr$^{-1}$ and 10$^{-3}$\msol{}~yr$^{-1}$, while the average accretion rates for less massive MYSOs are likely to be lower given the lifetimes we have obtained.

The luminosity functions of MYSOs and \HII{} regions presented in this paper can be used to test models of the time-evolution of the mass accretion rate and stellar properties. Such modelling will be discussed in a future publication (Davies \etal{} 2010, in prep.).

\acknowledgements 

The authors would like to thank the anonymous referee for comments which improved the quality and clarity of this paper. that JCM is partially funded by a Postgraduate Studentship and by a Postdoctoral Research Associate grant from the Science and Technologies Research Council of the United Kingdom (STFC). The authors would like to thank Cameron Reed for his help with the OB star luminosity function.


\begin{thebibliography}{49}
\expandafter\ifx\csname natexlab\endcsname\relax\def\natexlab#1{#1}\fi

\bibitem[{{Baker} \& {Menzel}(1938)}]{Baker1938}
{Baker}, J.~G., \& {Menzel}, D.~H. 1938, \apj, 88, 52

\bibitem[{{Behrend} \& {Maeder}(2001)}]{Behrend2001}
{Behrend}, R., \& {Maeder}, A. 2001, \aap, 373, 190

\bibitem[{{Boehm-Vitense}(1981)}]{Boehm-Vitense1981}
{Boehm-Vitense}, E. 1981, \araa, 19, 295

\bibitem[{{Bronfman} {et~al.}(2000){Bronfman}, {Casassus}, {May}, \&
  {Nyman}}]{Bronfman2000}
{Bronfman}, L., {Casassus}, S., {May}, J., \& {Nyman}, L. 2000, \aap, 358, 521

\bibitem[{{Bunn} {et~al.}(1995){Bunn}, {Hoare}, \& {Drew}}]{Bunn1995}
{Bunn}, J.~C., {Hoare}, M.~G., \& {Drew}, J.~E. 1995, \mnras, 272, 346

\bibitem[{{Chambers} {et~al.}(2009){Chambers}, {Jackson}, {Rathborne}, \&
  {Simon}}]{Chambers2009}
{Chambers}, E.~T., {Jackson}, J.~M., {Rathborne}, J.~M., \& {Simon}, R. 2009,
  \apjs, 181, 360

\bibitem[{{Chini} {et~al.}(1987){Chini}, {Kruegel}, \& {Wargau}}]{Chini1987}
{Chini}, R., {Kruegel}, E., \& {Wargau}, W. 1987, \aap, 181, 378

\bibitem[{{Churchwell} {et~al.}(2009){Churchwell}, {Babler}, {Meade},
  {Whitney}, {Benjamin}, {Indebetouw}, {Cyganowski}, {Robitaille}, {Povich},
  {Watson}, \& {Bracker}}]{Churchwell2009}
{Churchwell}, E., {et~al.} 2009, \pasp, 121, 213

\bibitem[{{Clarke} {et~al.}(2006){Clarke}, {Lumsden}, {Oudmaijer}, {Busfield},
  {Hoare}, {Moore}, {Sheret}, \& {Urquhart}}]{Clarke2006}
{Clarke}, A.~J., {Lumsden}, S.~L., {Oudmaijer}, R.~D., {Busfield}, A.~L.,
  {Hoare}, M.~G., {Moore}, T.~J.~T., {Sheret}, T.~L., \& {Urquhart}, J.~S.
  2006, \aap, 457, 183

\bibitem[{{Egan} {et~al.}(2003){Egan}, {Price}, \& {Kraemer}}]{Egan2003b}
{Egan}, M.~P., {Price}, S.~D., \& {Kraemer}, K.~E. 2003, \baas, 35, 1301

\bibitem[{{Egan} {et~al.}(1999){Egan}, {Price}, {Moshir}, {Cohen}, \&
  {Tedesco}}]{Egan1999}
{Egan}, M.~P., {Price}, S.~D., {Moshir}, M.~M., {Cohen}, M., \& {Tedesco}, E.
  1999, NASA STI/Recon Technical Report, 14854

\bibitem[{{Ellingsen}(2006)}]{Ellingsen2006}
{Ellingsen}, S.~P. 2006, \apj, 638, 241

\bibitem[{{Habing} \& {Israel}(1979)}]{Habing1979}
{Habing}, H.~J., \& {Israel}, F.~P. 1979, \araa, 17, 345

\bibitem[{{Henning} {et~al.}(1984){Henning}, {Friedemann}, {G{\"u}rtler}, \&
  {Dorschner}}]{Henning1984}
{Henning}, T., {Friedemann}, C., {G{\"u}rtler}, J., \& {Dorschner}, J. 1984,
  \an, 305, 67

\bibitem[{{Henning} {et~al.}(1990){Henning}, {Pfau}, \&
  {Altenhoff}}]{Henning1990}
{Henning}, T., {Pfau}, W., \& {Altenhoff}, W.~J. 1990, \aap, 227, 542

\bibitem[{{Hoare} \& {Franco}(2007)}]{Hoare2007b}
{Hoare}, M.~G., \& {Franco}, J. 2007, in Diffuse Matter from Star Forming
  Regions to Active Galaxies - A Volume Honouring John Dyson, ed. T.~W.
  {Hartquist}, J.~M. {Pittard}, \& S.~A.~E.~G. {Falle} (Springer), 61

\bibitem[{{Hosokawa} \& {Omukai}(2009)}]{Hosokawa2009}
{Hosokawa}, T., \& {Omukai}, K. 2009, \apj, 691, 823

\bibitem[{{Hosokawa} {et~al.}(2010){Hosokawa}, {Yorke}, \&
  {Omukai}}]{Hosokawa2010}
{Hosokawa}, T., {Yorke}, H.~W., \& {Omukai}, K. 2010, \apj, 721, 478

\bibitem[{{Hummer} \& {Storey}(1987)}]{Hummer1987}
{Hummer}, D.~G., \& {Storey}, P.~J. 1987, \mnras, 224, 801

\bibitem[{{Keto}(2003)}]{Keto2003}
{Keto}, E. 2003, \apj, 599, 1196

\bibitem[{{Lanz} \& {Hubeny}(2007)}]{Lanz2007}
{Lanz}, T., \& {Hubeny}, I. 2007, \apjs, 169, 83

\bibitem[{{Lumsden} {et~al.}(2002){Lumsden}, {Hoare}, {Oudmaijer}, \&
  {Richards}}]{Lumsden2002}
{Lumsden}, S.~L., {Hoare}, M.~G., {Oudmaijer}, R.~D., \& {Richards}, D. 2002,
  \mnras, 336, 621

\bibitem[{{Marshall} {et~al.}(2006){Marshall}, {Robin}, {Reyl{\'e}},
  {Schultheis}, \& {Picaud}}]{Marshall2006}
{Marshall}, D.~J., {Robin}, A.~C., {Reyl{\'e}}, C., {Schultheis}, M., \&
  {Picaud}, S. 2006, \aap, 453, 635

\bibitem[{{Martins} {et~al.}(2005){Martins}, {Schaerer}, \&
  {Hillier}}]{Martins2005}
{Martins}, F., {Schaerer}, D., \& {Hillier}, D.~J. 2005, \aap, 436, 1049

\bibitem[{{Meynet} \& {Maeder}(2003)}]{Meynet2003}
{Meynet}, G., \& {Maeder}, A. 2003, \aap, 404, 975

\bibitem[{{Molinari} {et~al.}(1996){Molinari}, {Brand}, {Cesaroni}, \&
  {Palla}}]{Molinari1996}
{Molinari}, S., {Brand}, J., {Cesaroni}, R., \& {Palla}, F. 1996, \aap, 308,
  573

\bibitem[{{Mottram} {et~al.}(2010){Mottram}, {Hoare}, {Lumsden}, {Oudmaijer},
  {Urquhart}, {Meade}, {Moore}, \& {Stead}}]{Mottram2010}
{Mottram}, J.~C., {Hoare}, M.~G., {Lumsden}, S.~L., {Oudmaijer}, R.~D.,
  {Urquhart}, J.~S., {Meade}, M.~R., {Moore}, T.~J.~T., \& {Stead}, J.~J. 2010,
  \aap, 510, A89

\bibitem[{{Mottram} {et~al.}(2007){Mottram}, {Hoare}, {Lumsden}, {Oudmaijer},
  {Urquhart}, {Sheret}, {Clarke}, \& {Allsopp}}]{Mottram2007}
{Mottram}, J.~C., {Hoare}, M.~G., {Lumsden}, S.~L., {Oudmaijer}, R.~D.,
  {Urquhart}, J.~S., {Sheret}, T.~L., {Clarke}, A.~J., \& {Allsopp}, J. 2007,
  \aap, 476, 1019

\bibitem[{{Mottram} {et~al.}(2011){Mottram}, {Hoare}, {Urquhart}, {Lumsden},
  {Oudmaijer}, {Robitaille}, {Moore}, {Davies}, \& {Stead}}]{Mottram2011}
{Mottram}, J.~C., {et~al.} 2011, \aap, 525, A149

\bibitem[{{Osterbrock}(1989)}]{Osterbrock1989}
{Osterbrock}, D.~E. 1989, {Astrophysics of gaseous nebulae and active galactic
  nuclei}, ed. {Osterbrock, D.~E.}

\bibitem[{{Paladini} {et~al.}(2004){Paladini}, {Davies}, \& {De
  Zotti}}]{Paladini2004}
{Paladini}, R., {Davies}, R.~D., \& {De Zotti}, G. 2004, \mnras, 347, 237

\bibitem[{{Palla} {et~al.}(1991){Palla}, {Brand}, {Comoretto}, {Felli}, \&
  {Cesaroni}}]{Palla1991}
{Palla}, F., {Brand}, J., {Comoretto}, G., {Felli}, M., \& {Cesaroni}, R. 1991,
  \aap, 246, 249

\bibitem[{{Palla} \& {Stahler}(1993)}]{Palla1993}
{Palla}, F., \& {Stahler}, S.~W. 1993, \apj, 418, 414

\bibitem[{{Reed}(2005)}]{Reed2005}
{Reed}, B.~C. 2005, \aj, 130, 1652

\bibitem[{{Robin} {et~al.}(2003){Robin}, {Reyl{\'e}}, {Derri{\`e}re}, \&
  {Picaud}}]{Robin2003}
{Robin}, A.~C., {Reyl{\'e}}, C., {Derri{\`e}re}, S., \& {Picaud}, S. 2003,
  \aap, 409, 523

\bibitem[{{Robitaille} {et~al.}(2007){Robitaille}, {Whitney}, {Indebetouw}, \&
  {Wood}}]{Robitaille2007a}
{Robitaille}, T.~P., {Whitney}, B.~A., {Indebetouw}, R., \& {Wood}, K. 2007,
  \apjs, 169, 328

\bibitem[{{Shu} {et~al.}(1987){Shu}, {Adams}, \& {Lizano}}]{Shu1987}
{Shu}, F.~H., {Adams}, F.~C., \& {Lizano}, S. 1987, \araa, 25, 23

\bibitem[{{Sridharan} {et~al.}(2002){Sridharan}, {Beuther}, {Schilke},
  {Menten}, \& {Wyrowski}}]{Sridharan2002}
{Sridharan}, T.~K., {Beuther}, H., {Schilke}, P., {Menten}, K.~M., \&
  {Wyrowski}, F. 2002, \apj, 566, 931

\bibitem[{{Tan} \& {McKee}(2003)}]{Tan2003}
{Tan}, J.~C., \& {McKee}, C.~F. 2003, in IAU Symposium, Vol. 221, Star
  Formation at High Angular Resolution, ed. M.~G. {Burton}, R.~{Jayawardhana},
  \& T.~L. {Bourke}, 274

\bibitem[{{Urquhart} {et~al.}(2007{\natexlab{a}}){Urquhart}, {Busfield},
  {Hoare}, {Lumsden}, {Clarke}, {Moore}, {Mottram}, \&
  {Oudmaijer}}]{Urquhart2007a}
{Urquhart}, J.~S., {Busfield}, A.~L., {Hoare}, M.~G., {Lumsden}, S.~L.,
  {Clarke}, A.~J., {Moore}, T.~J.~T., {Mottram}, J.~C., \& {Oudmaijer}, R.~D.
  2007{\natexlab{a}}, \aap, 461, 11

\bibitem[{{Urquhart} {et~al.}(2007{\natexlab{b}}){Urquhart}, {Busfield},
  {Hoare}, {Lumsden}, {Oudmaijer}, {Moore}, {Gibb}, {Purcell}, {Burton}, \&
  {Marechal}}]{Urquhart2007c}
{Urquhart}, J.~S., {et~al.} 2007{\natexlab{b}}, \aap, 474, 891

\bibitem[{{Urquhart} {et~al.}(2008{\natexlab{a}}){Urquhart}, {Busfield},
  {Hoare}, {Lumsden}, {Oudmaijer}, {Moore}, {Gibb}, {Purcell}, {Burton},
  {Mar{\'e}chal}, {Jiang}, \& {Wang}}]{Urquhart2008a}
---. 2008{\natexlab{a}}, \aap, 487, 253

\bibitem[{{Urquhart} {et~al.}(2008{\natexlab{b}}){Urquhart}, {Hoare},
  {Lumsden}, {Oudmaijer}, \& {Moore}}]{Urquhart2008b}
{Urquhart}, J.~S., {Hoare}, M.~G., {Lumsden}, S.~L., {Oudmaijer}, R.~D., \&
  {Moore}, T.~J.~T. 2008{\natexlab{b}}, in Astronomical Society of the Pacific
  Conference Series, Vol. 387, Massive Star Formation: Observations Confront
  Theory, ed. H.~{Beuther}, H.~{Linz}, \& T.~{Henning}, 381

\bibitem[{{Urquhart} {et~al.}(2009){Urquhart}, {Hoare}, {Purcell}, {Lumsden},
  {Oudmaijer}, {Moore}, {Busfield}, {Mottram}, \& {Davies}}]{Urquhart2009a}
---. 2009, \aap, 501, 539

\bibitem[{{Urquhart} {et~al.}(2011){Urquhart}, {Moore}, {Hoare}, {Lumsden},
  {Oudmaijer}, {Rathborne}, {Mottram}, {Davies}, \& {Stead}}]{Urquhart2011}
---. 2011, \mnras, 410, 1237

\bibitem[{{Walmsley}(1995)}]{Walmsley1995}
{Walmsley}, M. 1995, in Revista Mexicana de Astronomia y Astrofisica Conference
  Series, Vol.~1, Revista Mexicana de Astronomia y Astrofisica Conference
  Series, ed. S.~{Lizano} \& J.~M. {Torrelles}, 137

\bibitem[{{Wood} \& {Churchwell}(1989)}]{Wood1989b}
{Wood}, D.~O.~S., \& {Churchwell}, E. 1989, \apj, 340, 265

\bibitem[{{Wynn-Williams}(1982)}]{Wynn-Williams1982}
{Wynn-Williams}, C.~G. 1982, \araa, 20, 587

\bibitem[{{Yorke} \& {Bodenheimer}(2008)}]{Yorke2008}
{Yorke}, H.~W., \& {Bodenheimer}, P. 2008, in Astronomical Society of the
  Pacific Conference Series, Vol. 387, Massive Star Formation: Observations
  Confront Theory, ed. H.~{Beuther}, H.~{Linz}, \& T.~{Henning}, 189

\end{thebibliography}


\clearpage

\end{document}